%% file: output.tex
\documentclass[english,showpacs,preprint,superscriptaddress]{revtex4-1}

 \UseRawInputEncoding  
 
\usepackage{babel}
\usepackage[T1]{fontenc}      % 强制使用 Type 1 字体（兼容性更好）, 文本加粗
\usepackage{cmap}             % 确保 PDF 文本可搜索

\usepackage{mathrsfs}         % 花体， \mathcal \to \mathscr
\usepackage{bm}
\usepackage{color, xcolor}   % 文字高亮
\usepackage{ulem}
 
\usepackage{CJKutf8}  % 强制 UTF-8 支持

\usepackage{tipa}        % 国际音标 ， 希腊字母
\usepackage{lmodern}     % Latin Modern 字体（默认替代 Computer Modern）
\usepackage{amsmath,amsfonts,amssymb}
\usepackage{esint}     % 额外积分符号, 必须在`amsthm` 之前
\usepackage{extarrows}

\usepackage[ruled,linesnumbered]{algorithm2e}
\usepackage{algpseudocode}
\usepackage{graphicx}   % 插图
\usepackage{float}
\usepackage{tikz}       % 流程图
\usepackage{palatino}
\usetikzlibrary{arrows,shapes,chains}

\usepackage{subfig}     % 子图字体、索引等
\usepackage{booktabs}  % % 三线表
\graphicspath{figure/}% 配置图片的默认目录

\usepackage[justification=centering]{caption}
\usepackage{siunitx}

\usepackage{indentfirst}
\usepackage{setspace}

\makeatletter

\usepackage[T1]{fontenc}      % 文本加粗

\usepackage[title]{appendix}
\usepackage{hyperref}  % 创建超文本链接和PDF书签
\hypersetup{
    breaklinks = true,
    colorlinks = true,
    citecolor = {blue},
    urlcolor = {blue},
    linkcolor = {blue}
} 

\usepackage{cleveref}

\input{macros}


\input{macro_VFP}

\input{macro_VFP2}

\makeatother

\begin{document}

\title{Transport theory in moderately anisotropic plasmas: I, Collisionless aspects of axisymmetric velocity space}

% gbsn 对应宋体，gkai 为楷体，bsmi 为新宋体（繁体）
\author{Yanpeng Wang \begin{CJK}{UTF8}{gkai} (王彦鹏)\end{CJK}}
% \author{Yanpeng Wang}
\email{Email: tangwang@mail.ustc.edu.cn}
\affiliation{Department of Engineering Physics, Tsinghua University, Beijing, 100084, China}
\affiliation{School of Nuclear Sciences and Technology, University of Science and Technology of China, Hefei, 230026, China}

\begin{abstract} 
   
  A novel transport theory, based on the finitely distinguishable independent features (FDIF) hypothesis, is presented for scenarios when velocity space exhibits axisymmetry. In this theory, the transport equations are derived from the 1D-2V Vlasov equation, employing the spherical harmonics expansions (SHE) together with the King function expansion (KFE) in velocity space. The characteristic parameter equations (CPEs) are provided based on the general King mixture model (GKMM), serving as the constraint equations of the transport equations. It is a nature process to present the closure relations of transport equations based on SHE and KFE, successfully providing a kinetic moment-closed model (KMCM). This model is typically a nonlinear system, effective for moderately anisotropic non-equilibrium plasmas. 

   
\end{abstract}

\maketitle

 \UseRawInputEncoding


% \renewcommand{\cftdotsep}{\cftnodots}
% \cftpagenumbersoff{figure}
% \cftpagenumbersoff{table} 

% \setlength{\parskip}{0pt}

\begin{spacing}{0.35}  

\noindent

\section{Introduction}
\label{Introduction}   

Non-equilibrium statistical physics\cite{Dilip2014Nonequilibrium} constitutes the most vigorous frontier research in contemporary statistical physics and holds significant applications in plasma physics. One of the core issues pertains to how to precisely and effectively describe the collective behavior of charged particles within electromagnetic fields. 
The Vlasov equation\cite{Vlasov1968} serves as the cornerstone for describing the evolution of collisionless plasmas. Nevertheless, the high-dimensional {and nonlinear} nature of the Vlasov equation
renders its direct solution computationally costly\cite{Johnston1960, Tzoufras2011, Thomas2012, Taitano2015I}. To surmount these challenges, researchers have developed various dimensionality reduction approaches, {such as moment approaches\cite{Grad1949, Chapman1953, Braginskii1958, Hunana2022, Cai2015AFramework}, gyrokinetic method\cite{Lee1987, Brizard2007}, discrete Boltzmann method\cite{zhang2019, zhu2023} (DBM)}. Among those methods, the moment approaches{, including the Grad's method\cite{Grad1949} and the Braginskii models\cite{Braginskii1958, Hunana2022},} are preferred due to its balance between physical intuitiveness and computational efficiency. {Additionally, high-order moment convergence (similar to the conservation laws) is also an advantage of the moment approaches\cite{wang2024higherorder}, which is still a challenge for the method of directly solving the Vlasov equation with conservation constraints\cite{Taitano2015Amass}.}

% The solution of this problem directly determines the accuracy and scope of application of the moment method. 
The success of the moment method hinges on a key issue\cite{Schunk1977, Cai2021}: how to introduce a reasonable closure hypothesis into the higher-order moment equation to truncate the infinitely dimensional moment hierarchy. 
The 13-moment method proposed by Grad\cite{Grad1949} was undoubtedly a milestone, based on the near-equilibrium assumption. Grad's framework expands the distribution function (DF) into a Hermite polynomial\cite{Arfken1971} and achieves closure by simply truncating higher-order terms. However, its capacity to describe strongly non-equilibrium plasmas is limited\cite{Schunk1977}. 
Additionally, the Grad's method introduces mathematically non-rigorous assumptions in the truncation of higher order moments, which may further limit its range of adaptation. {Similarly, the closure of Braginskii model\cite{Braginskii1958}, which is based on the Chapman-Enskog expansion\cite{Chapman1953}, also poses a challenge\cite{Hunana2022}. Moreover, its general form\cite{Hunana2022}, as a general moment theory, is highly complex with a set of formulations.} 

In order to overcome the drawbacks of the moment method, researchers have proposed a variety of improvement schemes\cite{Cai2021, Sarna2020, Rangan2007, Li2021, Dimarco2010}, {including the hyperbolic moment method\cite{Cai2014Globally}, regularized moment method\cite{Cai2014Globally}, maximum entropy method\cite{Levermore1996}, quadrature-based moment method\cite{Fox2009Higher-order} and entropic quadrature closure method\cite{Bohmer2020}.} 
Nevertheless, these methods still fail to fundamentally address the universality of the closure hypothesis. {Moreover, both Grad's method and Braginskii models, which are founded on the near-equilibrium assumption, fail into the category of the perturbation theory\cite{Eu1998TheChapman-Enskog}. For highly nonlinear problems, especially those with multiple extremes\cite{Tomasz2020II}, the applicability of perturbation theory is still an open question in mathematics.}

% Under strong non-equilibrium conditions such as fusion plasmas and turbulent transport, the DF may deviate significantly from the local Maxwellian distribution, leading to the invalidity of Grad's closure hypothesis. 

% For example, the extended moments method improves the accuracy of the model by increasing the number of higher-order moments, but this often comes at the expense of computational efficiency and is difficult to overcome the limitations of the local thermodynamic equilibrium assumption. Another idea is to introduce non-polynomial basis functions to better adapt to the non-Gaussian properties of DF. 

Within this context, the \textit{finitely distinguishable independent features} (FDIF) hypothesis has been introduced in previous works\cite{wang2024Relaxationmodel, wang2024Aconservative} {, to replace the traditional near-equilibrium assumption and offer a nonlinear framework for solving the Vlasov-Fokker-Planck (VFP) equation}. The central idea is that although the DF might possess a complex structure in the phase space, its evolution can be dominated by a finite number of key characteristics. These characteristics can not only capture the main kinetic behavior of the DF but also reflect its higher-order statistical properties to a certain extent. 
Based on the FDIF hypothesis, a novel framework\cite{wang2024higherorder}, including a meshfree approach\cite{wang2024Aconservative} and a moment approach\cite{wang2024Relaxationmodel, wang2024general}, is provided for addressing the nonlinear simulation in fusion plasmas. In both of these approaches, the independent features are certain geometric properties of the DF in phase space and are typically nonlinear functions of the kinetic moments\cite{wang2024Relaxationmodel}. 

In this {nonlinear} framework {which is not a perturbation theory}, the {spherical harmonics expansion}\cite{Johnston1960, Bell2006, Tzoufras2011} (SHE) is employed for the angular coordinate in velocity space, and the \textit{King function expansion}\cite{wang2024Relaxationmodel, wang2024Aconservative} (KFE) method is employed for the speed coordinate. {The theoretical convergence of KFE has been demonstrated\cite{wang2024Relaxationmodel} with the aid of Wiener's Tauberian theorem\cite{Wiener1932}. In the meshfree approach\cite{wang2024Aconservative} when investigating the collision aspects with axisymmetric velocity space, KFE} has been demonstrated to be a moment convergent method {and the convergence order can reach up to 16. In the moment approach,} a nonlinear relaxation model is presented for a homogeneous plasmas in scenario with shell-less spherically symmetric velocity space\cite{wang2024Relaxationmodel}. Subsequently, the general relaxation model for spherically symmetric velocity space with shell structures\cite{wang2024Relaxationmodel, Min2015} is provided in reference\cite{wang2024general}. In this study, the transport theory
{focusing on the collisionless aspects when velocity space exhibits axisymmetry is provided,} based on the FDIF hypothesis for moderately anisotropic\cite{Bell2006, Tzoufras2011, wang2024Aconservative} non-equilibrium plasmas.
% (details in Sec. \ref{Closure in l space}).

The following sections of this paper are organized as follows. Sec. \ref{Maxwell-Vlasov system} provides an introduction to the Vlasov equation and its spectrum form. Sec. \ref{Transport equations} discusses the transport equations in cases where velocity space exhibits axisymmetry, including the closure relations for kinetic moment-closed model. Finally, a summary of our work is presented in Sec.\SEC{Conclusion}.

\end{spacing}

\begin{spacing}{0.7}

\section{Maxwell-Vlasov system}
\label{Maxwell-Vlasov system}

The evolution of distribution function (DF) in collisionless fusion plasmas can be described by the Vlasov\cite{Vlasov1968} equation. For species $a$, denoting the DF as $f=f \left(\r, \v,t \right)$ in the phase space $\left(\r, \v \right)$, where $\r$ and $\v$ denote the physics space and velocity space respectively. Thus, the Vlasov equation can be expressed as
  \begin{eqnarray}
      \ddt{f} + \v \cdot \nabla {f} + \frac{Z_a}{m_a} \left(\En + \v \times \Bn \right) \cdot \ddbfv f &=& 0 ~.\label{Vlasov}
  \end{eqnarray}
Here, $\nabla$ and $\nabla_{\v}$ respectively represent the gradient operator of the physics space and the velocity space. Symbols $Z_a$ and $m_a$ respectively denote the charge number and mass of species $a$. In accordance with the convention and to avoid symbol confusion, the unit of $Z_a$ in this article is the unit charge $e$. Otherwise, the International System of Units is adopted.

The electric field intensity, $\bsE$, and magnetic flux density, $\bsB$, in Eq.\EQ{Vlasov} can be described by the Maxwell's equations
  \begin{eqnarray}
      \nabla \times \En &=& - \ddt \Bn, \label{dtB} 
      \\
      \nabla \times \frac{\Bn}{\mu_0} &=& \varepsilon_0 \ddt \En + \Jn_q, \label{dtE} 
      \\
      \nabla \cdot \En &=& {\rho_q}/{\varepsilon_0}, \label{DE} 
      \\
      \nabla \cdot \Bn &=& 0 , \label{DB}
  \end{eqnarray}
where $\varepsilon_0$ and $\mu_0$ respectively denote the dielectric constant and the magnetic permeability of vacuum. Symbols $\rho_q$ and $\Jn_q$ respectively represent the charge density and electric current density of the plasmas system, as defined in Sec. \ref{Velocity moments}. The above equations\EQ{Vlasov}-\EQo{DB} constitute the {Maxwell-Vlasov} (MV) system. In this content, the relativistic effect is ignored.

\subsection{Vlasov spectrum equation for axisymmetric velocity space}
\label{Vlasov spectrum equation}

When the velocity space is expressed in terms of spherical-polar coordinates $\v(v, \theta,\phi)$, the DF can be expanded by employing {spherical harmonic expansion}\cite{Johnston1960, Bell2006, Tzoufras2011, wang2024Relaxationmodel, wang2024Aconservative, wang2024general, wang2024higherorder} (SHE) method in velocity space. {Notes that the DF is assumed to be a smooth function in the velocity space.}
Moreover, when the velocity space exhibits {axisymmetry} with symmetry axis $\bse_z$, the azimuthal mode number $m$ is identically zero in the spectral space $(j,m)$. As a result, the collisionless plasmas system can be characterized by a one-dimensional, two-velocity (1D-2V) Vlasov spectral equation.

Before presenting the Vlasov spectral equation, the SHE for the DF of species $a$ can be expressed as follows
  \begin{eqnarray}
      f \left(\r,\v,t \right) &=& \sumloq \sumlnl \delta_m^0 \flm \left(\r, \rmv ,t \right) \Ylm \left(\mu, \phi \right) , \label{fshe}
  \end{eqnarray}
where {$\delta_m^0$ is the  the Kronecker symbol,} $l \in \bbN$ and $m \in \bbZ$ . Symbol $\Ylm$ denotes the complex form of spherical harmonic\cite{Arfken1971} without the normalization coefficient, $\Nlm = \sqrt{\frac{2l+1}{4\pi} \frac{(l-m)!}{(l+m)!}}$. 
By applying the inverse transformation to Eq.\EQ{fshe}, the $(l, m)^{th}$-order amplitude function can be obtained, which reads
   \begin{eqnarray}
       \flm \left(\r,\rmv,t \right) &=& \delta_m^0 \frac{1}{\left (\Nlm \right)^2} \int_{-1}^1 \int_0^{2 \pi} \left (\Ylmabs \right)^* f \left (\r, \v, t \right) \rmd \phi  \rmd \mu, \quad l \in \bbN, \quad m \ge 0,  \label{fl}
   \end{eqnarray}
where $(\cdots)^*$ represents the complex conjugation. Note that the order $m \equiv 0$ for the axisymmetric velocity space.  Thus, for simplicity, the superscripts of amplitudes can be omitted. For instance, $\fl$ will be employed instead of $\flo$.

In the Cartesian coordinate system of the physics space, $\r(x,y,z)$, the field vectors such as electric field intensity, magnetic flux density and gradient operator, $\nabla$, can be expressed as
   \begin{eqnarray}
       \bsE &=& \left [E_x \quad E_y \quad E_z \right],
       \quad 
       \bsB \ = \ \left [B_x \quad B_y \quad B_z \right],
       \quad
       \nabla \ = \ \left [\ddx \quad \ddy \quad \ddz \right]~. \label{D}
   \end{eqnarray}
Based on the SHE formula\EQ{fshe} of DF, the $l^{th}$-order \textit{Vlasov spectrum equation}\cite{Tzoufras2011} can be obtained, reads
   \begin{eqnarray}
       \ddt \fl \left(\r,\rmv,t \right) &=& \Al + \El + \Bl~.  \label{Vlasovfl}
   \end{eqnarray}
The anisotropic terms with $|m| \ge 1$ are identically zero.
The amplitude of the convection terms at the $l^{th}$-order, resulting from the non-uniform DF in physical space, is given by
  \begin{eqnarray}
      \Al \left(\r,\rmv,t \right)  & = &  - \rmv \ddz \fAl~.  \label{Al}
  \end{eqnarray}
Similarly, the $l^{th}$-order amplitude of effect terms induced by the macroscopic electric field and magnetic field can be expressed as follows
  \begin{eqnarray}
      \El \left(\r,\rmv,t \right) &=& \left( {Z_a}/{m_a} \right) E_z  \fEl,  \label{El}
      \quad
      \Bl \left(\r,\rmv,t \right) \ = \ \left( {Z_a}/{m_a} \right) \left (i B_z  \fBl \right)~.  \label{Bl}
  \end{eqnarray}
The symbol $i$ denotes the imaginary unit. The terms related to other components of electric field intensity and magnetic flux density, namely $E_x$, $E_y$, $B_x$ and $B_y$, are all exactly zero for scenarios with axisymmetric velocity space. 
  
Symbols $\fAl$, $\fEl$ and $\fBl$ in Eqs.\EQ{Al}-\EQo{Bl} are functions of amplitude function, reads
   \begin{eqnarray}
      \fAl \left(\r,\rmv,t \right) &=& \left [\CAln \ \ \CAlp \right] \cdot  \left [\fln \ \ \flp \right],  \label{\fAl}
      \\
      \fEl \left(\r,\rmv,t \right) &=& \left [\CEIln \ \ \CEln \ \ \CEIlp \ \ \CElp \right] \cdot \left [\ddrmv \fln \ \ \frac{\fln}{\rmv} \ \ \ddrmv \flp \ \ \frac{\flp}{\rmv} \right],  \label{fEl}
      \\
      \fBl \left(\r,\rmv,t \right) &=& \CBl \fl~.  \label{fBl}
   \end{eqnarray}
{
Operator $\bsr \cdot \bsE$ denotes the dot product of vectors $\bsr$ and $\bsE$, given by
   \begin{eqnarray}
       \r \cdot \bsE  &=& [x \quad y \quad z] \cdot [E_x \quad E_y \quad E_z] \ = \ x E_x + y E_y + z E_z  ~.
   \end{eqnarray}
}
The scalar coefficients in Eqs.\EQ{\fAl}-\EQo{fBl} such as $\CAln$ are functions of $l$. Among these, the coefficients associated with the spatial convection terms are
\begin{eqnarray}
    \CAln \ = \ \frac{l}{2l-1},  \quad \CAlp \ = \ \frac{l+1}{2l+3} ~.
\end{eqnarray}
The coefficients associated with the electric field effect terms and the magnetic field effect terms are
\begin{eqnarray}
    \CEln \ = \ \CAln,  
    \quad 
    \CElp \ = \ \CAlp,
    \quad 
    \CBl \ = \ - m \ \equiv \ 0 
\end{eqnarray}
and
\begin{eqnarray}
    \CEIln \ = \ - (l-1) \CEln,
    \quad
    \CEIlp  \ = \ (l+2) \CElp ~.
\end{eqnarray}
Therefore, the magnetic effect term, $\Bl$, remains constantly zero due to the fact that $\CBl$ is zero when velocity space exhibits axisymmetry. However, $B_z$ may not be zero. 
{It is a constant field without gradient and does not vary with time. This can be verified in the corresponding Maxwell's equations\EQ{dtB}-\EQo{DB} in Cartesian coordinate system $\bsr (x,y,z)$, which can be reformulated as
\begin{eqnarray}
\frac{\partial E_z}{\partial y} &=&  
\frac{\partial E_z}{\partial x} \ = \
\frac{\partial B_z}{\partial t} \ = \ 
\frac{\partial B_z}{\partial y} \ = \ 
\frac{\partial B_z}{\partial x} \ = \
\frac{\partial B_z}{\partial z} \ = \ 0 ,  \label{DBz} 
\\
\epsilon_0 \frac{\partial E_z}{\partial t} &=& - \bse_z \cdot {\bsJ_q},  \label{dtEz} 
% \end{eqnarray}
% and
% \begin{eqnarray}
\\
 \frac{\partial E_z}{\partial z} &=& \frac{\rho_q}{\epsilon_0} ~. \label{dzEz} 
\end{eqnarray}
This indicates that the scenario featuring axisymmetric velocity space is associated with the situation where an uniform constant magnetic field exists in the presence of a uniform parallel electric field.
}

Therefore, when velocity space exhibits axisymmetry with axis $\bse_z$, all the gradient fields in $x$ and $y$ direction are zero, such as $\ddx f_l (\r,v,t)$ and $\ddy f_l (\r,v,t)$. 
Additionally, when the velocity space exhibits spherical symmetry, all amplitudes with order $l \ge 1$ will be zero, that is $f_{l \ge 1} \equiv 0$. Moreover, coefficients with order $l-1$ will be zero, that is $\CMAln = \CEln = \CEIln \equiv 0$. Hence, the Vlasov spectrum equation\EQ{Vlasovfl} will be reduced to
   \begin{eqnarray}
       \ddt \fl \left(\r,\rmv,t \right) &\equiv& 0, \quad \forall l~.  \label{Vlasovf0}
   \end{eqnarray}
This indicates that when the velocity space exhibits spherical symmetry, the DF of collisionless plasmas will remain unchanged over time. {Subsequently, all the components of electric field and magnetic field will be zero.}

\subsection{Weak form of Vlasov spectrum equation}
\label{Weak form of Vlasov spectrum equation}

Weak form is typically more useful, especially for numerical computation. 
One can directly present the weak form in velocity space based on Eq.\EQ{Vlasovfl}. 
Multiplying both sides of the Vlasov spectral equation\EQ{Vlasovfl} by $4 \pi m_a \rmv^{j+2} \rmd \rmv$ and integrating over the semi-infinite interval $\rmv = \left [0, \infty \right)$, simplifying the result gives the $(j,l)^{th}$-order Vlasov spectral equation in weak form, reads
\begin{eqnarray}
    \ddt \left[ 4 \pi m_a \int_0^\infty \rmv^{j+2} \fl \rmd \rmv \right] &=&
    4 \pi m_a \int_0^\infty \rmv^{j+2} \left(\Al + \El + \Bl \right) \rmd \rmv, \quad \forall l  , \label{Vlasovflweak}
\end{eqnarray}
where $j \ge - l -2$ and functions $\Al$, $\El$ and $\Bl$ are provided by Eq.\EQ{Al}-\EQo{Bl} respectively.

\section{Transport equations}
\label{Transport equations}

Before presenting the transport equations for moderately anisotropic plasmas (the definition is given in Sec. \ref{Kinetic moment-closed model}), the definition of \textit{kinetic moment} introduced in reference\cite{wang2024Relaxationmodel, wang2024Aconservative} will be offered. It is a functional of amplitude of the DF, given by
\begin{equation}
    \calMjl \left(\r, t \right) \ = \ \calM_j \left[\fl \right] \ = \ 4 \pi m_a \int_0^{\infty} (\rmv)^{j+2} \fl \left(\r,\rmv,t \right) \rmd \rmv, \quad \forall (j,l)   ~. \label{Mjl}
\end{equation}
Moreover, the traditional velocity moments\cite{Freidberg2014} and their relationships relative to the above kinetic moments are provided in Sec.\SEC{Velocity moments}
.
      
\subsection{Kinetic moment evolution equation}
\label{Kinetic moment evolution equation}

By utilizing the definition of kinetic moment\EQ{Mjl}, the weak form of the $(j,l)^{th}$-order Vlasov spectral equation\EQ{Vlasovflweak} can be reformulated as the moment evolution equation. The vector form of the $(j,l)^{th}$-order \textit{kinetic moment evolution equation} (KMEE) can be obtained, which is presented as follows
\begin{eqnarray}
    \ddt \calMjl \left(\r, t \right) &=& - \nabla \cdot \calMjplI 
     + \frac{Z_a}{m_a} \bsE \cdot \calMjnlI, \quad \forall (j,l)  
     , \label{KMEEjl0}
\end{eqnarray}
where the symbol $l \pm$ indicates that the corresponding function is relative to the kinetic moments of order $l \pm1$. The vectors $\calMjplI$ and $\calMjnlI$ can be expressed as
  \begin{eqnarray}
      \calMjplI \left(\r,t \right) &=&  
      \begin{bmatrix}
          0 & 0  & \CvMAzjl \cdot \calMzjplI
      \end{bmatrix}  , \label{MAjl}
      \\ 
      \calMjnlI \left(\r,t \right) &=&  
      \begin{bmatrix}
          0 & 0  & \CvMEzjl \cdot \calMzjnlI
      \end{bmatrix} ~. \label{MEjl}
      % \\ 
      % \calMjl \left(\r,t \right) &=&  
      % \begin{bmatrix}
      %     0 & 0  & \CBl \cdot \calMzjl
      % \end{bmatrix} ~. \label{MBjl}
  \end{eqnarray}
   
The vectors associated to the spatial convection terms and electric field effect terms in Eq.\EQ{MAjl}-\EQo{MEjl} are
  \begin{eqnarray}
      \calMzjplI \left(\r,t \right) &=& \left [\calMjpln \ \ \calMjplp \right] , \label{calMAzjplIm}
      \\
      \calMzjnlI \left(\r,t \right) &=& \left [\calMjnln \ \ \calMjnlp \right] ~. \label{calMEzjnlIm}
      % \\
      % \calMzjl \left(\r,t \right) &=& i \calMjl, \label{calMBzjl}
  \end{eqnarray}
The corresponding coefficients are functions of $j$ and $l$, which can be given by
  \begin{eqnarray}
      \CvMAzjl &=& \left [\CMAln \ \ \CMAlp \right],  \label{CvMAzjl} 
      \quad
      \CvMEzjl \ = \ \left [\CMEjln \ \ \CMEjlp \right],  \label{CvMEzjl} 
      % \\
      % \CvBzjl &=& \CBl \ \equiv \ 0,  \label{CvBzjl} 
  \end{eqnarray}
where
\begin{eqnarray}
    \CMEjln \ = & (j+l+1) \CEln,
    \quad
    \CMEjlp \ = & (j-l) \CElp~.
\end{eqnarray}

The KMEE\EQ{KMEEjl0} represents the general form of \textit{transport equation} when the velocity space exhibits axisymmetry. It will be a high-precision approximate theory to the original 1D-2V Vlasov equation\EQ{Vlasov} by choosing an appropriate collection of order $(j,l)$, especially for moderately anisotropic plasmas (details in Sec.\ref{Kinetic moment-closed model}). The KMEE\EQ{KMEEjl0} can also be reformulated as
\begin{eqnarray}
    \ddt \calMjl \left(z, t \right) &=& - \CvMAzjl \cdot \ddz \calMzjplI
     + \frac{Z_a}{m_a} E_z \CvMEzjl \cdot \calMzjnlI 
     \label{KMEEjl0scalar} 
\end{eqnarray}
or in a more direct and concrete form
\begin{eqnarray}
\begin{aligned}
    \ddt \calM_{j,l} \left(z, t \right) =& - \ddz \left [\frac{l}{2l-1} \calM_{j+1,l-1} + \frac{l+1}{2l+3} \calM_{j+1,l+1} \right]
    \\
     &+ \frac{Z_a}{m_a} E_z \left [(j+l+1) \frac{l}{2l-1} \calM_{j-1,l-1} + (j-l) \frac{l+1}{2l+3} \calM_{j-1,l+1} \right]  ~. \label{KMEEjl0scalarCoef} 
\end{aligned}
\end{eqnarray}
{Field $E_z$ of plasmas with species number $N_s$ satisfies Eqs.\EQ{dtEz}-\EQo{dzEz}, and it can be rewritten with the utilization of Eqs.\EQ{M0}-\EQo{M1} as follows:
\begin{eqnarray}
    \frac{\partial E_z}{\partial t} &=& - \frac{1}{3} \frac{1}{\epsilon_0} \sum_a \frac{Z_a}{m_a} \calM_{1,1},  \label{dtEzMjl} 
    \\
     \frac{\partial E_z}{\partial z} &=& \frac{1}{\epsilon_0} \sum_a \frac{Z_a}{m_a} \calM_{0,0}~. \label{dzEzMjl} 
\end{eqnarray}
}
Moreover, the explicit equations of conservation laws of mass, momentum and energy are presented in Sec.\SEC{Conservation laws}. It should be noted that the kinetic effects are described by the higher orders of kinetic moments\EQ{Mjl}.

Specially, when the velocity space exhibits spherical symmetry, all kinetic moments with order $l \ge 1$ will be zero due to $f_{l \ge 1} \equiv 0$. Note that coefficients with order $l-1$ will be zero in this situation. Therefore, the KMEE\EQ{KMEEjl0} will be reduced to
\begin{eqnarray}
    \ddt \calMjl \left(z, t \right) &\equiv& 0, \quad \forall (j,l)  
     ~. \label{KMEEj00}
\end{eqnarray}
This indicates that the field effect terms, including the spatial convection, electric field and magnetic field effects terms, all vanish in scenarios with spherically symmetric velocity space. In other words, if the collision effects are disregarded in scenarios with spherically symmetric velocity space, the macro state of the plasmas system does not change over time. Otherwise, with collision effects, the corresponding form of KMEE is provided in reference\cite{wang2024Relaxationmodel, wang2024general}.

\subsection{Velocity moments and conservation laws}
\label{Velocity moments and conservation laws}

The kinetic moment represented by Eq.\EQ{Mjl} possesses a concise and unified definition, offering a beautiful form of KMEE\EQ{KMEEjl0} derived from the Vlasov equation\EQ{Vlasov}. {The arbitrary-order equation of transport theory represented by KMEE has a unified form. This makes the transport theory normative and relatively simple when compared with the general Braginskii model\cite{Hunana2022}.} However, the physical pictures are not as evident as that of the traditional moment equations\cite{Freidberg2014}. In this section, the relationship between these two descriptions will be illustrated. 

\subsubsection{Velocity moments}
\label{Velocity moments}

 KMEE\EQ{KMEEjl0} indicates that the kinetic moments of orders $(j,l)$ represent the \textit{intensive quantities} and of orders $(j+1,l \pm 1)$ denote the \textit{fluxes} in the plasmas system, while those of orders $(j-1,l \pm 1)$ stand for the quantities associated with the electric field. The first few orders of them represent the components of the traditional \textit{velocity moment}\cite{Johnston1960}, denoted by the symbol $<\cdots,f> = m_a \int_{\v} (\cdots) f (\r, \v, t) \rmd \v$.
 
Similar to Tzoufras\cite{Tzoufras2011}, velocity moment can be expressed in terms of the amplitude $\fl$. When velocity space exhibits axisymmetry, the first few orders can be expressed as
  \begin{eqnarray}
      \left < \rmv^j, f \right> &=& \calMjo, \label{Mg0}
      \\
      \left < \rmv^j \frac{\v}{\rmv}, f \right> &=&  \frac{1}{3} \calM_j \left[\bsf_1 \right], \label{Mg1}
      \\
      \left < \rmv^j \frac{\v \v}{\rmv^2}, f \right> &=&  \frac{1}{3} \calMjo \olra{I} + \frac{2}{3 \times 5} \calM_j \left[\bsf_2 \right], \label{Mg2}
  \end{eqnarray}
where $\v \v$ denotes a second-order tensor, symbol $\olra{I}$ represents the unit tensor and
  \begin{eqnarray}
      \bsf_1  &=&  \left [0 \quad 0 \quad f_1 \right],  \label{fj1I}
      \quad 
      \bsf_2  \ = \ \frac{1}{2} \scrT_{2,2} f_2 ~. \label{fj2I}
  \end{eqnarray}
Matrix $\scrT_{2,2}$ is defined as
   \begin{eqnarray}
       \scrT_{2,2}  & = & 
       \begin{bmatrix}
          - 1 &   0  & 0 \\
           0  & - 1  & 0 \\
           0  &  0   & 2 \\
      \end{bmatrix}~. \label{T20}
   \end{eqnarray}
Applying the definition of kinetic moment and relations\EQ{Mg0}-\EQo{Mg2}, one can directly acquire the first few orders of velocity moments\cite{Freidberg2014}. The mass density (zeroth-order), momentum (first-order) and the total press tensor (second-order) can be expressed as
   \begin{eqnarray}
       \rho_a \left(z, t \right) & = & \left <1,f \right > \ = \ \calM_{0,0} , \label{M0}
       \\
       \Ia \left(z, t \right) \ & = & \left <\v,f \right > = \ \Iza \bse_z \ = \ \frac{1}{3} \calM_{1,1} \bse_z,  \label{M1}
       \\
       \olra{P_a} \left(z, t \right) & = & \left <\v \v,f \right > \ = \   \frac{1}{3} \calM_{2,0} \olra{I} + \frac{1}{3 \times 5} \scrT_{2,2} \calM_{2,2} ~. \label{M2}
   \end{eqnarray}
   
Similar to Freidberg\cite{Freidberg2014}, the scalar pressure is defined as follows
   \begin{eqnarray}
        p_a \left(z, t \right) &=& \frac{1}{3} m_a \left < w_a^2, f \right >, \label{pa}
   \end{eqnarray}
where the random thermal motion, $\wn_a = \v - \ua$ and let $w_a=|\wn_a|$. Similarly, the temperature can be defined as a functional of the DF, as given by
      \begin{eqnarray}
        T_a \left(z,t \right) &=& \frac{2}{3} \frac{m_a}{n_a} \left < \frac{w_a^2}{2}, f \right > ~. \label{Ta}
      \end{eqnarray}
The scalar pressure, number density $n_a=\int_{\v} f\rmd \v$ and average velocity $\ua = n_a^{-1} \int_{\v} \v f \rmd \v$ satisfy the following relations
   \begin{eqnarray}
       p_a \left(z, t \right) & = & n_a T_a,
       \quad
       n_a \left(z, t \right) \ = \ \rho_a / m_a , \label{na}
       \quad
       \ua \left(z, t \right) \ = \ \Iza \bse_z / \rho_a ~. \label{ua}
   \end{eqnarray}
The total energy, $K_a = (m_a / 2) \int_{\v} |\v|^2 f \rmd \v$, can be calculated as follows
   \begin{eqnarray}
       K_a \left(z, t \right) & = & \calM_{2,0} / 2 ~. \label{Ka}
   \end{eqnarray}
Similarly, the charge density $\rho_q$ and current density $\bsJ_q$ of plasmas can be expressed as
   \begin{eqnarray}
       \rho_q \left(z, t \right) & = & \sum_a Z_a n_a, \label{rhoq}
       \quad 
       \bsJ_q \left(z, t \right) \ = \ \sum_a Z_a n_a \ua  ~. \label{Jq}
   \end{eqnarray}

Let $u_a=|\ua|$, applying the following definition of inner energy, 
    \begin{eqnarray}
      \epsilon_a \left(z,t \right)  &=& \frac{3}{2} n_a T_a \ = \  K_a  -  E_{k_a}  ,  \label{epsilona}
    \end{eqnarray}
   % \begin{eqnarray}
   %     K_a \left(z, t \right) & = & \frac{3}{2} n_a T_a +  \frac{1}{2} \rho_a u_a^2 ~.
   % \end{eqnarray}
where the kinetic energy,
    \begin{eqnarray}
      E_{k_a} \left(z,t \right)  &=&  \rho_a u_a^2 / 2, \label{Eka}
    \end{eqnarray}
one can obtain the following relation
  \begin{eqnarray}
      T_a \left(z,t \right)  &=& \frac{2}{3}  \frac{K_a}{n_a}  -  \frac{2}{3} \times \left ( \frac{1}{2} m_a u_a^2 \right)  ~.  \label{TaM}
  \end{eqnarray}
The total press tensor also can be expressed as
\begin{eqnarray}
       \olra{P_a} & = & \left (p_a + \frac{1}{3} \rho_a u_a^2 \right) \olra{I} + \frac{1}{3 \times 5} \scrT_{2,2} \calM_{2,2}~. \label{PaI2}
\end{eqnarray}
The thermal velocity is defined as follows
      \begin{eqnarray}
        \vath \left(z,t \right) &=& \sqrt{{2 T_a} / {m_a}} ~. \label{vath0}
      \end{eqnarray}
Obvious, it is a function of $\rho_a$\EQ{M0}, $I_a$\EQ{M1} and $K_a$\EQ{Ka}, reads
\begin{eqnarray}
    \vath (t) &=& \sqrt{\frac{2}{3} \left(\frac{2 K_a}{\rho_a} - \left(\frac{I_a}{\rho_a} \right)^2 \right)} ~.\label{vath}
\end{eqnarray}

The intrinsic pressure tensor and the viscosity tensor (anisotropic part of the intrinsic pressure tensor) are respectively defined as
  \begin{eqnarray}
      \olra{p_a} \left(z,t \right) &=& m_a \left < \wn_a \wn_a, f \right > \ = \ \olra{P_a} - \rho_a \ua \ua , \label{paI} \\
      \olra{\Pi}_a \left(z,t \right) &=& \olra{p_a} - p_a \olraI
      % \ = \ \frac{1}{3 \times 5} \scrT_{2,2} \calM_{2,2}
      ~. \label{PiaI} 
  \end{eqnarray}
The intrinsic pressure tensor and the viscosity tensor possess symmetry, that is
  \begin{eqnarray}
      p_{a_{ij}} \left(z,t \right) &=& m_a \left < w_{a_i} w_{a_j}, f \right > = p_{a_{ji}}, \label{paij}
      \\
      \Pi_{a_{ij}} \left(z,t \right) &=& p_{a_{ij}} - p_a \delta_i^j = \Pi_{a_{ji}}
      ~. \label{Piaij}
  \end{eqnarray}
The intrinsic heat flux vector due to random motion and the total energy flux caused by random motion respectively are
  \begin{eqnarray}
      \qn_a \left(z,t \right) &=& ({m_a} / {2})  \left <  w_a^2 \wn_a,  f \right > = q_{a_z} \bse_z, \label{bfqa}
      \\
      \Qn_a \left(z,t \right) &=& \frac{m_a}{2}  \left <  \rmv^2 \v, f \right > \ = \ \frac{1}{2 \times 3} \calM_3 \left[\bsf_1 \right]
      ~. \label{bfQa}
      % \\
      % Q_{a_z} \left(z,t \right) &=& \left(K_a +  p_{a_{zz}} \right)  \uza  +  q_{a_z}
      % , \label{bfQaz}
  \end{eqnarray}
Moreover, one can obtain the following relation
  % \begin{eqnarray}
  %     Q_{a_z} \left(z,t \right) &=&  K_a  \uza +  \ua \cdot  \olra{p_a} + q_{a_z} ~.
  % \end{eqnarray}
  % \begin{eqnarray}
  %     Q_{a_z} \left(z,t \right) &=&  K_a  \uza +  \ua \cdot  \olra{P_a} - \rho_a \uza^3 + q_{a_z} ~.
  % \end{eqnarray}
  % \begin{eqnarray}
  %     Q_{a_z} \left(z,t \right) &=&  K_a  \uza + \frac{1}{3} \uza \calM_{2,0} + \frac{2}{3 \times 5}  \uza \calM_{2,2} - \rho_a \uza^3 + q_{a_z} ~.
  % \end{eqnarray}
  \begin{eqnarray}
      Q_{a_z} \left(z,t \right) &=& \frac{1}{2 \times 3} \calM_{3,1} = \frac{5}{3} \uza K_a + \frac{2}{3 \times 5}  \uza \calM_{2,2} - \rho_a \uza^3 + q_{a_z} ~. \label{bfQaz}
  \end{eqnarray}
Utilizing the definitions of above quantities, one can readily derive the conservation laws of mass, momentum and energy, which will provided in next section.

% Particularly, when $j=l$,
% \begin{eqnarray}
%     \ddt \calM_{l,l} \left(z, t \right) &=& - \ddz \left [\frac{l}{2l-1} \calM_{l+1,l-1} + \frac{l+1}{2l+3} \calM_{l+1,l+1} \right]
%      + l \frac{2l+1}{2l-1} \frac{Z_a}{m_a} E_z \calM_{l-1,l-1}  ~. \label{KMEEll0scalarCoef} 
% \end{eqnarray}
% When $l=0$,
% \begin{eqnarray}
%     \ddt \calMjo \left(z, t \right) &=& - \frac{1}{3} \ddz \calM_{j+1,1}
%      + \frac{j}{3} \frac{Z_a}{m_a} E_z \calM_{j-1,1} ~. \label{KMEEj00scalarC} 
% \end{eqnarray}
% When $l=1$,
% \begin{eqnarray}
%     \ddt \calM_{j,1} \left(z, t \right) = - \ddz \left (\calM_{j+1,0} + \frac{2}{5} \calM_{j+1,2} \right)
%     + \frac{Z_a}{m_a} E_z \left [(j+2) \calM_{j-1,0} + \frac{2}{5} (j-1) \calM_{j-1,2} \right]  ~. \label{KMEEj10scalarCoef} 
% \end{eqnarray}

\subsubsection{Conservation laws}
\label{Conservation laws}

When $(j,l)=(0,0)$, $(j,l)=(1,1)$ and $(j,l)=(2,0)$, the KMEE\EQ{KMEEjl0} is simplified to the \textit{mass, momentum and energy conservation laws} when velocity space exhibits axisymmetry, which can be directly derived from Eq.\EQ{KMEEjl0scalarCoef},
% \begin{eqnarray}
% \begin{aligned}
%     \ddt \calM_{j,1} \left(z, t \right) =& - \ddz \left [\CMAln \calMjpln + \CMAlp \calMjplp \right]
%     \\
%      &+ \frac{Z_a}{m_a} E_z \left [\CMEln \calMjnln + \CMElp \calMjnlp \right]  ~. \label{KMEEj10scalar} 
% \end{aligned}
% \end{eqnarray}
\begin{eqnarray}
    \ddt \calMoo \left(z, t \right) &=& - \frac{1}{3} \ddz \calM_{1,1}, \label{KMEE000scalar} 
    \\
    \ddt \calM_{1,1} \left(z, t \right) &=& - \ddz \left (\calM_{2,0} + \frac{2}{5} \calM_{2,2} \right)
    + 3 \frac{Z_a}{m_a} E_z \calMoo , \label{KMEE110scalar} 
    \\
    \ddt \calM_{2,0} \left(z, t \right) &=& - \frac{1}{3} \ddz \calM_{3,1}
     + \frac{2}{3} \frac{Z_a}{m_a} E_z \calM_{1,1} ~.  \label{KMEE200scalar} 
\end{eqnarray}
Employing the definitions of velocity moments in Sec. \ref{Velocity moments}, the conservation laws represented by Eq.\EQ{KMEE000scalar}-\EQo{KMEE200scalar} can be reformulated as
\begin{eqnarray}
    \ddt \rho_a \left(z, t \right) &=& - \ddz \Iza, \label{Cn} 
    \\
    \ddt \Iza \left(z, t \right) &=& - \nabla \cdot \olra{P_a} + \frac{Z_a}{m_a} \rho_a E_z , \label{CI} 
    \\
    \ddt K_a \left(z, t \right) &=& -  \ddz Q_{a_z} 
     + \frac{Z_a}{m_a}  \Iza E_z ~.  \label{CK} 
\end{eqnarray}
The terms on the right side of Eq.\EQ{CK} respectively depict the net energy flux flowing into the fluid element surface and the power of work performed by the electric field on the fluid element (Ohmic heating power) of species $a$.
Eq.\EQ{Cn}-\EQo{CK} are the traditional form of mass, momentum and energy conservation for collisionless plasmas when the velocity space exhibits axisymmetry. 

% \begin{eqnarray}
%     \ddt \calM_{1,1} \left(z, t \right) &=& - \ddz \left (\calM_{2,0} + \frac{2}{5} \calM_{2,2} \right) + 3 \frac{Z_a}{m_a} \calMoo E_z  ~.  
%     \\
%     \ddt \Iza \left(z, t \right) &=& - \frac{1}{3} \ddz \left (2 K_a + \frac{2}{5} \calM_{2,2} \right)
%     + \frac{Z_a}{m_a} E_z \calMoo ,
%     \\
%     \ddt \Iza \left(z, t \right) &=& - \ddz \left (p_a + \frac{1}{3} \rho_a u_a^2 + \frac{2}{3 \times 5} \calM_{2,2} \right)
%     + \frac{Z_a}{m_a} \rho_a E_z
% \end{eqnarray}

% \begin{eqnarray}
%     \frac{1}{2} \ddt \calM_{2,0} \left(z, t \right) &=& - \frac{1}{2} \frac{1}{3} \ddz \calM_{3,1}
%      + \frac{1}{3} \frac{Z_a}{m_a} E_z \calM_{1,1} ~. 
% \end{eqnarray}

Additionally, employing the relations of $\Iza$\EQ{ua}, $\olra{P_a}$\EQ{PaI2} and $Q_{a_z}$\EQ{bfQaz}, the equation of continuity can be expressed as
  \begin{eqnarray}
      \ddt \rho_a + \ddz  \left (\rho_a \uza \right )  \ = \ 0 \label{Cnmhd}
  \end{eqnarray}
and equation of motion for fluid elements will be
  \begin{equation}
      \rho_a \DDt \uza \ = \ - \nabla \cdot \olra{\Pi}_a - \ddz p_a + Z_a n_a E_z~. \label{CImhd} 
  \end{equation}
{
The material derivative operator $\rmd / \rmd t$ for species $a$ is
  \begin{eqnarray}
      \DDt &=& \ddt + \ua \cdot \nabla ~. \label{DDt} 
  \end{eqnarray}
}
The terms on the right side respectively represent viscous force, thermal pressure and electric field force. By making use of the equations of mass\EQ{Cn} and momentum\EQ{CI} conservation, one can also acquire the heat balance equation
  \begin{equation}
      \frac{3}{2} n_a \DDt T_a \ = \ - \left(\olra{p_a} \cdot \nabla \right) \cdot \ua  - \ddz q_{a_z}~. \label{CKmhd} 
  \end{equation}
The items on the right hand denote the power of work done by viscous force (internal friction force) and heat conduction.
Eq.\EQ{Cnmhd}-\EQo{CKmhd} share the same form as that of traditional two-fluid equations\cite{Freidberg2014}. Nevertheless, those equations are typically not a closed system in the general situation of fusion plasmas. The closure relations for moderately anisotropic plasmas will be provided in next section.

\subsection{Kinetic moment-closed model for moderately anisotropic plasmas}
\label{Kinetic moment-closed model}

Unfortunately, in general scenarios, the transport equations described by KMEE\EQ{KMEEjl0} with a truncated order of $l$ and finite collection of order $j$ typically lack closure, as indicated in the $(j,l)^{th}$-order equation which contains moments of higher order such as those with $j+1$ or $l+1$. Instead of the traditional near-equilibrium assumption employed by Grad\cite{Grad1949}, a \textit{finitely distinguishable independent features} (FDIF) hypothesis\cite{wang2024Relaxationmodel, wang2024Aconservative} can be adopted to enclose the aforementioned transport equations\EQ{KMEEjl0}. This hypothesis states that a fully ionized plasmas system has a finite number of distinguishable independent characteristics\cite{wang2024Relaxationmodel}. Under the FDIF hypothesis, the transport equations can be enclosed in the $(j,l)$ space, especially for moderately anisotropic plasmas\cite{wang2024Aconservative}.

\subsubsection{Closure in $(l)$ space}
\label{Closure in l space}

For moderately anisotropic plasmas\cite{Bell2006, wang2024Aconservative}, the series on the right side of the Eq.\EQ{fshe} will converge rapidly. 
Similar to Grad's method\cite{Grad1949}, the order $l$ can be truncated at a maximum value, $l_M$, where the subscript $M$ stands for $max$, satisfying that $\max|\fhl(\r,\rmv,t)|\le Atol_{df}$ where $\fhl = \vath^3 n_a^{-1} \fl$. Here, $Atol_{df}$ is a tolerance with a default value of $10^{-14}$. Thus, $0 \le l \le l_M$. 
    
  \begin{figure}[htp]
	\begin{center}
		\includegraphics[width=0.8\linewidth]{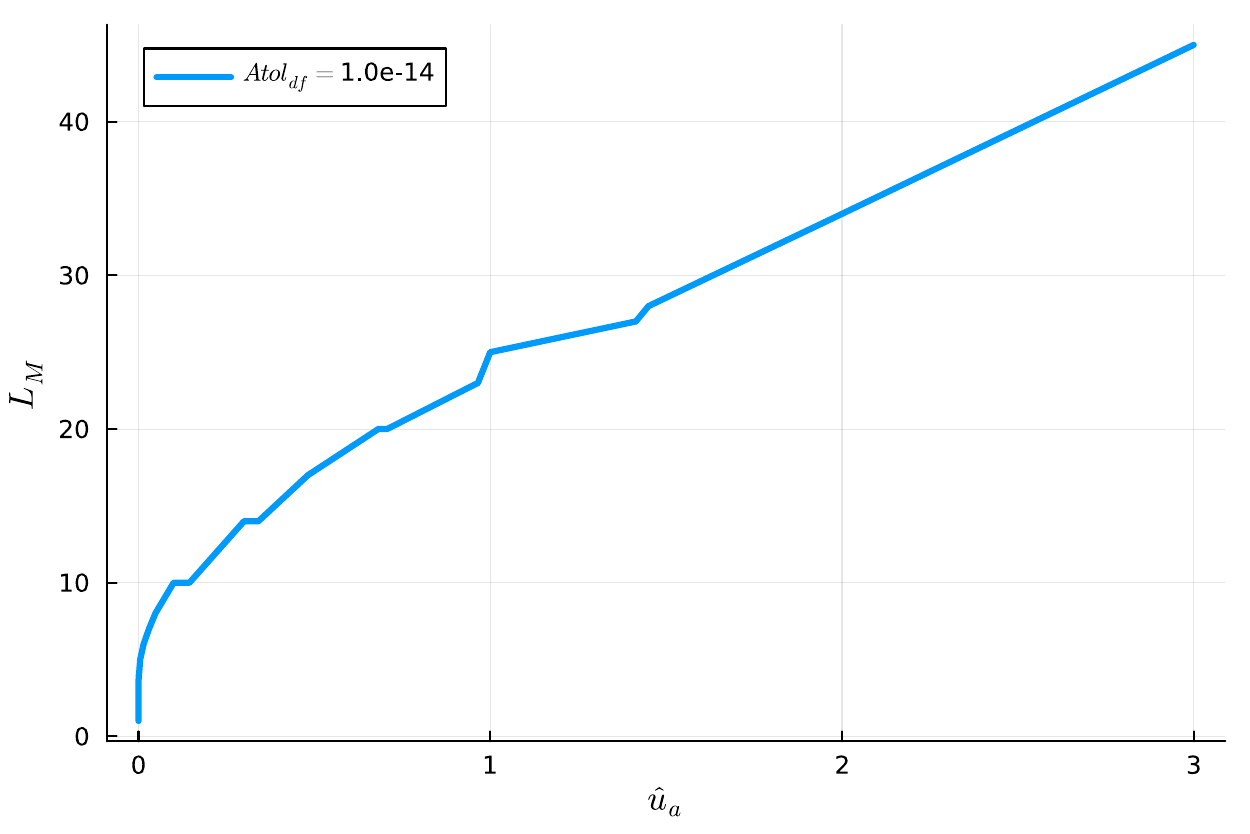}
	\end{center}
	\caption{Convergence of SHE for drift-Maxwellian distribution: Truncated order $l_M$ as a function of $\hat{u}_a$ when tolerance $Atol_{df}=10^{-14}$.}
	\label{FigLMuh}
  \end{figure}

The convergence of SHE for drift-Maxwellian distribution is depicted in Fig.\FIG{FigLMuh}. It indicates that the truncated order $l_M$ is a monotonic function of normalized average velocity, $\uzha=\uza / \vath$. For example with $Atol_{df}=10^{-14}$, the maximum order is $l_M = 14$ when $\uzha = 0.3$, while $l_M = 45$ when $\uzha = 3$. \textit{Moderately anisotropic plasmas} can be defined as the system without significant asymmetric characteristics such as beam, where the truncated order $l_M \le 45$ when tolerance $Atol_{df}=10^{-14}$. A more rigorous definition of \textit{anisotropy} is provided in Sec. \ref{Closure in j space} with the closure in $(j)$ space.
Based on the above definition, $l_M$ is typically not larger than $45$ for moderately anisotropic plasmas. This method is referred to as the \textit{natural truncation method} in this article.

When the natural truncation method is applied in the $(l)$ space and the truncated order $l_M$ is sufficiently large (perhaps varying with time), the amplitude functions with order larger than $l_M$ can simply be negligible. Subsequently, the higher-order kinetic moments with order $ l \ge l_M+1$ in the KMEE\EQ{KMEEjl0} are negligible small quantities, that is
  \begin{eqnarray}
      \calM_{j,l} \left(z, t \right) & \equiv & 0, \quad \forall j, \quad l \ge l_M + 1 ~.  \label{NCR}
  \end{eqnarray}
This is the \textit{natural closure relation} (NCR) in the $(l)$ space. By employing NCR, the KMEE represented by Eq.\EQ{KMEEjl0scalar} with order $l_M \ge 1$ can be simplified to
\begin{eqnarray}
    \ddt \calM_{j,l_M} \left(z, t \right) &=& - \CMAln \ddz \calMjpln
     + \CMEln \frac{Z_a}{m_a} E_z \calMjnln, \quad l = l_M  ~. \label{KMEEjlM0scalar} 
\end{eqnarray}

Note that when $l=0$, the coefficients $\CMAln = \CMEjln = \CEln \equiv 0$. Hence, KMEE\EQ{KMEEjl0scalar} reduces to
\begin{eqnarray}
    \ddt \calMjo \left(z, t \right) &=& - \CMAlp \ddz \calMjplp
     + \CMElp \frac{Z_a}{m_a} E_z \calMjnlp, \quad l = 0 ~. \label{KMEEj00scalar} 
\end{eqnarray}
It can be observed that the KMEE with order $(j, l=0)$ is independent of the kinetic moments of order $l-1$. That is to say that when $l_M=0$, applying NCR\EQ{NCR} results in
\begin{eqnarray}
    \ddt \calMjo \left(z, t \right) &=& 0, \quad \forall j ~. \label{KMEEj00lM0} 
\end{eqnarray}
That is Eq.\EQ{KMEEj00}, representing the situation for collisionless plasmas when the velocity space exhibits spherical symmetry.

\subsubsection{Closure in $(j)$ space}
\label{Closure in j space}

For moderately anisotropic plasmas, the FDIF hypothesis\cite{wang2024Relaxationmodel, wang2024Aconservative} is an effective assumption to offer closure and capture the nonlinearity of the plasmas system. According to this hypothesis,  it is assumed that the $l^{th}$-order amplitude distribution has distinguishable independent features with a number of $N_l$. Then, if we know the values of $N_l$ kinetic moments $\calMjl$ with a collection of $(j)$, these features can be determined by these given kinetic moments. Here, the collection of $(j)$ is referred to as a vector. For example,  $\bsj_l = [j_m, \cdots, j-dj, j, j+dj, \cdots, j_M]$, where $dj \in \bbN^+$ and subscripts $m$ and $M$ stand for $min$ and $max$ respectively. Subsequently, $\calM_{j\le j_m-1,l}$ and  $\calM_{j\ge j_M+1,l}$ are dependent on the kinetic moments within collection $\bsj_l$. Therefore, the {closure relation in $(j)$ space} can be expressed as
  \begin{eqnarray}
      \calM_{j,l} \left(z, t \right) &=& \calM_{j,l} \left(\calM_{j_m,l}, \calM_{j_m+dj,l}, \cdots, \calM_{j_M-dj,l}, \calM_{j_M,l} \right), \quad j \notin \bsj_l ~. \label{Mjclose}
  \end{eqnarray}
The concrete formula of the above equation can be derived from the following model based on the FDIF hypothesis.

To capture the nonlinearity of the amplitude function $\fl$ (nature for the general situation of fusion plasmas), a \textit{King function expansion} (KFE) method is introduced in Ref.\cite{wang2024Relaxationmodel, wang2024Aconservative}, as given by
    \begin{eqnarray}
        \fl \left(z, \rmv / \vath,t \right) &=& \frac{2 \pi} {\pi^{3/2}} \navth \sum_{s=1}^{\NKa}  \left [ \nhasl \Kl \left(\rmvh;\uzhasl,\vhathsl \right) \right] ~.  \label{KFE}
    \end{eqnarray}
Here, $\NKa \in \bbN^+$, which maybe vary with different $l$. The normalized speed coordinate, $\rmvh = \rmv / \vath$. Symbols
$\nhasl$, $\uzhasl$ and  $\vhathsl$ represent the characteristic parameters of $s^{th}$ sub-distribution of $\fl$, respectively referred to as weight, group velocity and group thermal velocity. Generally, the characteristic parameters typically depend on $l$ in common scenarios. Function $\Kl$ denotes a new special function introduced in Ref.\cite{wang2024Aconservative}, which is associated with {function $Besseli$ (} the first class of modified Bessel function {)},
    \begin{eqnarray}
        \Kl \left(\rmvh;\iota,\sigma \right) &=& \frac{(l+1/2)}{\sigma^2 \sqrt{2 \left |\iota \right | \rmvh}} \left(\frac{\left |\iota \right |}{\iota} \right)^l \e^{-\sigma^{-2} \left (\rmvh^2 + \iota^2  \right)} \mathrm{Besseli} \left(\frac{2l+1}{2}, 2 \frac{\left |\iota \right |}{\sigma^2} \rmvh \right) ~.  \label{King}
    \end{eqnarray}
Moreover, the reference\cite{wang2024Aconservative} has shown that KFE is a moment convergent method. Here, Eq.\EQ{KFE} is referred to as the \textit{general King mixture model} (GKMM) and the responding plasmas state referred to as \textit{local sub-equilibrium}\cite{wang2024higherorder}. Specially, when the characteristic parameters are independent of $l$, the GKMM reduces to \textit{King mixture model\cite{wang2024Aconservative}} (KMM) and the responding plasmas state is referred to as \textit{local quasi-equilibrium}\cite{wang2024higherorder}.

Substituting Eq.\EQ{KFE} into the definition of kinetic moment\EQ{Mjl} yields the \textit{characteristic parameter equations} (CPEs) for arbitrary  order $(j,l)$ with $j \ge - l - 2$.
{
For the convenience of academic communication, let
      \begin{eqnarray}
            \calD_{j,l} \left(z, t \right) &=& \frac{\calMjl}{{C_M}_j^l \rho_a \left (\vath \right )^j},
      \end{eqnarray}
where coefficient
      \begin{eqnarray}
            {C_M}_j^l &=& \frac{1}{2^{(j-l)/2}}  \frac{(l + j + 1)!!}{(2 l - 1)!!} ~.
      \end{eqnarray} 
% gbsn 对应宋体，gkai 为楷体，bsmi 为新宋体（繁体）
% Function $\calD_{j,l}$ is referred to as the $(j,l)^{th}$-order \textit{dientropy} (International Phonetic Alphabet (IPA): /\textit{dai'\textepsilon ntr\textschwa pi}/), a term derived from Chinese character `\CJKfontspec{黑体}\selectfont{焍}' (Pinyin: `\textit{d\`{i}'}, IPA: /\textit{ti}\textdownstep/).
The $\calD$ function, $\calD_{j,l}$, is referred to as the $(j,l)^{th}$-order \textit{dientropy} (International Phonetic Alphabet (IPA): /\textit{dai'\textepsilon ntr\textschwa pi}/), a term derived from Chinese character `\begin{CJK}{UTF8}{bsmi} 焍 \end{CJK}' (Pinyin: `\textit{d\`{i}'}, IPA: /\textit{ti}\textdownstep/).
% The function $\mathcal{D}_{j,l}$ is defined as the $(j,l)^{\text{th}}$-order \textit{dientropy} 
% (phonetically rendered as \textipa{/daI"\textteshligntr@pi/}), 
% a term derived from the Chinese character `焍' 
% (pinyin: \textit{d\`{i}}, IPA: \textipa{/ti\textsextoneshortlow/}).
% The function $\mathcal{D}_{j,l}$ is defined as the $(j,l)^{\text{th}}$-order \textit{dientropy} (pronounced as /daɪˈɛntɹəpi/), derived from Chinese `焍'  (\textit{dì}, /ti˥˩/).
Therefore, the CPEs can be expressed as
%       \begin{eqnarray}
%             \calMjl &=& 
%             \rho_a \left (\vath \right )^j 
%             \times \frac{2}{\sqrt{\pi }}  \Gamma \left(\frac{j+3}{2}\right)
%             \nhasl \left(\vhathsl \right)^j 
%             \IFI \left[-\frac{j}{2};\frac{3}{2};\left(\xiasl \right)^2\right] , \quad l =0
%             \label{CPEsL0}
%       \end{eqnarray} 
% and
      \begin{eqnarray}
            \calD_{j,l} \left(z, t \right) =
            \left \{ 
      \begin{aligned}
            &
            \frac{2}{\sqrt{\pi }} \frac{1}{{C_M}_j^l} \Gamma \left(\frac{j+3}{2}\right)
            \nhasl \left(\vhathsl \right)^j 
            \IFI \left[-\frac{j}{2};\frac{3}{2};\left(\xiasl \right)^2\right] , \quad \quad \quad \quad \quad \quad \quad \quad \quad l =0,
            \\ &  \left \{ 
            \begin{aligned}
            &
            (-1)^{l+1}  \sum_{s=1}^{\NKa} C_{j,l}^{0,s}
            \left \{
            \IFI \left[j_{l_3};\frac{1}{2};\left(\xiasl \right)^2\right]
            -\IFI \left[j_{l_3};\frac{3}{2};\left(\xiasl \right)^2\right]
            \right. \\ &  \left.
            \quad + \sum_{k=1}^{\lfloor l/2 \rfloor} c_{l,k} \frac{(j-l+1+2 k)!!}{(j-l+1)!!} \left(\xiasl \right)^{2 k} 
            % \right. \\ &  \left.
            \left[(l-2 k) \IFI \left(j_{l_3} +k;\frac{1}{2};\left(\xiasl \right)^2\right)
            \right. \right. \\ &  \left.  \left. \quad 
            -  (1+2 k) (l-k) \IFI \left(j_{l_3} +k;\frac{3}{2};\left(\xiasl \right)^2\right)\right]
            \right \} ,
            \quad \quad \quad \quad \quad \quad \quad \quad \quad \quad \quad l \ge 1,
            \end{aligned}
            \right. 
      \end{aligned}
      \right. 
      \label{CPEs}
      \end{eqnarray} 
where symbol $\IFI (a,b,z)$ represents the Kummer confluent hypergeometric $\IFI$ function.
Symbols $j_{l_3} = \left(j - l - 3 \right)/2$, $\xiasl = \uzhasl / \vhathsl$, $\lfloor x \rfloor$ means flooring $x$ to the integer and
      \begin{eqnarray}
            C_{j,l}^{0,s} &=& \frac{1}{\sqrt{\pi}} \frac{1}{{C_M}_j^l} \frac{(2 l + 1)!!}{2^{l-1}} \nhasl \left(\vhathsl \right)^{j}  \left(\xiasl \right)^{-l}  \Gamma \left(j_{l_3}\right) e^{- \left(\xiasl \right)^2} ~.
      \end{eqnarray} 
The coefficient $c_{l,k}$ in Eq.\EQ{CPEs} when $l \ge 1$ is a function only dependent on $l$ and $k$, and it can be expressed as
\begin{eqnarray}
    c_{l,k} = \frac{2}{3} \frac{1}{(2)_{k-1} (\frac{5}{2})_{k-1}} \frac{(l - 2 k + 1)_{k-1}}{\prod_{n=0}^{k-1} (2 l - 1 - 2n)}, \quad l \in \bbN^+, \quad k=1,2,\cdots, \lfloor l/2 \rfloor ~.
\end{eqnarray}
Here, the function $(x)_k$ represents the Pochhammer symbol related to $x$. 
}

Especially, when order $j \in \left\{(l+2j_p-2)|j_p \in \bbN^+ \right\}$, {it is easy to verify that} the CPEs represented by Eq.\EQ{CPEs} will be reduced as follows
      \begin{eqnarray}
            \calD_{j,l} \left(z, t \right) &=&
            \sum_{s=1}^{\NKa} \nhasl (\vhathsl)^j \left(\frac{\uzhasl}{\vhathsl} \right)^l 
            \left [1 + \sum_{\gamma=1}^{(j-l)/2} C_{j,l}^\gamma \left (\frac{\uzhasl}{\vhathsl} \right)^{2\gamma} \right] , \ \forall l ,   \label{CPEsd2}
      \end{eqnarray}
where coefficients $C_{j,l}^\gamma$ will be given by
      \begin{eqnarray}
          C_{j,l}^\gamma &=& 2^\gamma \frac{(2 l + 1)!!  C_\gamma^{(j-l)/2}}{(2 l + 2 \gamma + 1)!!}~.
      \end{eqnarray}
Symbol $C_\gamma^{(j-l)/2}$ represents the binomial coefficient. 
Hence, given a specific collection $\bsj_l$ with a number $N_l=3 \NKa$ and the values of the corresponding kinetic moments, the characteristic parameters can be determined by the well-posed CPEs\EQ{CPEs}. 

It is nature to offer the closure in $(j)$ space based on the CPEs.
{
For nonhomogeneous plasmas, the closure relation will be given based on Eq.\EQ{CPEs} with $dj=1$, which can be expressed as follows
      \begin{eqnarray}
            \calD_{j,l} \left(z, t \right) &=& \calD_{j,l} \left (j,l, \nhasl, \uzhasl, \vhathsl, \NKa \right)  , \quad j \notin \bsj_l, \quad \forall l ~.   \label{CCR}
      \end{eqnarray}
}
This equation is referred as the \textit{characteristic closure relation} (CCR) in the $(j)$ space. {Furthermore, for homogeneous plasmas without electric field\cite{wang2024Relaxationmodel, wang2024general, wang2024Aconservative}, a more effective form of CCR can be provided based on Eq.\EQ{CPEsd2} with $dj=2$,} given by
      \begin{eqnarray}
            \calD_{j,l} \left(z, t \right) &=& \sum_{s=1}^{\NKa} \nhasl (\vhathsl)^j \left(\frac{\uzhasl}{\vhathsl} \right)^l 
            \left [1 + \sum_{\gamma=1}^{(j-l)/2} C_{j,l}^\gamma \left (\frac{\uzhasl}{\vhathsl} \right)^{2\gamma} \right] , j \notin \bsj_l, \forall l ~.  \quad  \label{CCRd2}
      \end{eqnarray}

The definition of \textit{anisotropy} for plasmas can also be provided with the aid of characteristic parameters in KFE\EQ{KFE}, which is associated with the group velocity of $f_l$. For species $a$, it is defined that \textit{anisotropy} is equivalent to the maximum value of $\left |\uzhasl \right| / \vhathsl$ for all orders in $(s,l)$ space and is denoted as
      \begin{eqnarray}
            \fraka_a \left(z, t \right) &=&
            \max \left [\left ({\left |\uzhasl  \right|} / {\vhathsl}  \right)_{s \in [1,2,\cdots,N_l/3],l \in [0,1,2,\cdots,l_M]}  \right] ~.   \label{anisotropy}
      \end{eqnarray}
It is evident that the anisotropy of drift-Maxwellian plasmas will be
      \begin{eqnarray}
            \fraka_a \left(z, t \right) &=& |\uza|  / \vath~.    \label{anisotropyfDM}
      \end{eqnarray}
In particular, the anisotropy becomes zero when $\uza = 0$, corresponding to the Maxwellian distribution which is the isotropic part. Therefore, \textit{isotropy} is naturally a special state of anisotropy. 
Based on the definition of anisotropy, the \textit{weakly anisotropic plasmas} are defined as a system in which the anisotropy $\fraka_a \le 0.1$, including the system in the thermodynamic equilibrium state and the near-equilibrium state. Meanwhile, the \textit{moderately anisotropic plasmas} is defined as a system where the anisotropy $\fraka_a \le 3$, encompassing the weakly anisotropic plasmas and the so-called subsonic regions to low supersonic plasmas.

For moderately anisotropic plasmas, by selecting an appropriate collection $\bsj_l$ and the truncated order $l_M$, the transport equation presented by KMEE\EQ{KMEEjl0} in combination with constraint equations represented by CPEs\EQ{CPEs} and the closure relations described by NCR\EQ{NCR} and CCR\EQ{CCR} constitutes a \textit{kinetic moment-closed model} (KMCM).

\subsection{{Application of KMCM}}
\label{Application of KMCM}

{
The presented moment method, which solves the Vlasov equation based on SHE together with KFE framework, is transformed into addressing the nonlinear KMCM in the sense of moment convergence. Within the situation of moderately anisotropic plasma, KMCM controls the model error through reasonable truncation and approximates the Vlasov equation with specified accuracy. The efficiency and accuracy of this method lie in disregarding the high-order asymmetry of the velocity space and eliminating the additional high-frequency effect while maintaining the convergence of the selected moments. KMCM seems to be more complex than the traditional low-order moment theory, which is an inevitable outcome of retaining more asymmetry and high-frequency effects in the velocity space. It should be noted that it is more user-friendly than the general Braginskii model\cite{Hunana2022}. As a kinetic moment theory, its success lies in the following three aspects: 
\begin{spacing}{0.2}  
\begin{itemize}
    \item[(i)] 
    The core of the theory lies in the symmetry idea contained in SHE and the continuous smoothness of the King function.
    \item[(ii)] 
    The foundation of this system is established based on the strict mathematical proof in reference\cite{wang2024Relaxationmodel} that the KFE can approximate any one-dimensional continuous smooth function. The convergence can be accomplished based on the Wiener's Tauberian theorem \cite{Wiener1932}. The numerical results in reference\cite{wang2024Aconservative} demonstrate that the KFE has moment convergence, and the corresponding polynomial convergence order can reach 16 (approaching spectral convergence).
    \item[(iii)] 
    The numerical stability lies theoretically in that SHE and KFE are similar to low-pass filters in velocity space, playing the role of smoothing functions. As a kinetic theory, the key to the success of KMCM lies in that the continuous smoothness of KFE greatly alleviates the CFL condition while retaining the nonlinear effects in velocity space.
\end{itemize}
\end{spacing}
A simpler and more intuitive comprehension is that the King function is a special spline function, which is more efficient for the plasma DF in the speed coordinate. Its uniqueness lies in the following: The property that the derivatives of the $l^{th}$-order King function respective to $v$ is zero at the point $v = 0$ when derivative order is larger than $l$. Similarly, all derivatives of any order are zero at $v \to \infty$. These properties are consistent with the ones of $f_l$\cite{Tzoufras2011, wang2024Aconservative}. That is, the boundary conditions of the speed coordinate and their arbitrary derivatives naturally meet the characteristics of the plasmas system (naturally satisfy the boundary conditions of the velocity space). This property is one of the reasons for the rapid convergence of KFE, accounting for the numerical stability along with smoothness assumption of DF. 
}

\subsubsection{{Numerical solution}}
\label{Numerical solution} 

{
Mathematically, KMCM is a system of constrained first-order partial differential equations. Similar to solving the Vlasov equation and the traditional moment theory, KMCM can be addressed by constructing specific numerical algorithms of different levels. The numerical solution of KMCM is divided into two steps: the time integration of KMEE\EQ{KMEEjl0} and the solution of the constraint equation. With reasonable treatment of the spatial gradient term, time integration can employ the Rung-Kutta method \cite{Rackauckas2017}. The greatest challenge lies in the solution of CPEs\EQ{CPEs}: at each time step, the characteristic parameters are obtained from the kinetic moments (given by Rung-Kutta step). CPEs are nonlinear algebraic systems that can be constructed in under-determined/over-determined/well-posed forms (but none of them is suitable for direct conversion to linear algebraic systems). Based on the principle of simplicity, the well-posed CPEs is selected. 
}

{
The theory of general algebraic equations reveals that the analytical method is merely applicable to algebraic systems with polynomial order no more than 5, while the general solution lies in optimization algorithms. The construction of specific optimization algorithms is also classified into constrained/unconstrained forms (and various specific schemes). Reference \cite{wang2024Aconservative} provides several simple numerical algorithms, such as the L01jd2NK scheme which is only designed for near-equilibrium problem. As is well-known, initial value sensitivity is a key issue of optimization algorithms in multi-extremum problems. It is a difficult problem that needs to be overcome when extended to general algorithms. However, referring to PINN and other relevant research experiences, the general optimization algorithm required by KMCM can be further enhanced. In the future, research outcomes dedicated to general optimization algorithms will be released. 
}

\subsubsection{{Physical application}}
\label{Physical application}

{
KMCM can be strictly degenerated from the case with axisymmetric velocity space (1D-2V Vlasov) to the spherical symmetric case (0D-1V Vlasov) and the simplest thermodynamic equilibrium state. It can be further extended to the general velocity space case with nonlinear Fokker-Planck collision terms (3D-3V VFP). Physically, the application scope of KMCM can be analyzed from two levels of moment theory and kinetic theory:
\begin{spacing}{0.2}  
\begin{itemize}
    \item[(i)] 
    Moment level: KMCM can be regarded as an alternative to the Grad moment theory or the Braginskii model (including magnetohydrodynamics and two-fluid model). Hence, for problems where the traditional moment theory is applicable, KMCM is theoretically suitable as well. 
    \item[(ii)]  
    Kinetic level: KMCM uses high-order moments to describe kinetic effects. Within the category where the DF has continuous smoothness, the untruncated infinite-dimensional equation system of KMCM is equivalent to the description of the Vlasov equation. The truncated KMCM based on the FDIF hypothesis is a theory that effectively describes nonlinear effects. This theory is under the premise of ensuring the three major conservation laws and selecting high-order moments to satisfy uniform convergence (similar to conservation). In the category of moderately anisotropic plasmas, KMCM can approximate the Vlasov equation with specified accuracy. Therefore, within this parameter range, for problems where the Vlasov equation is applicable, KMCM is also theoretically suitable.
\end{itemize}
\end{spacing}
After the derivation of the Vlasov equation and prior to the optimal solution, KMCM is merely truncated based on fast convergence without any linearization. Hence, KMCM is applicable for the study of plasma problems featuring highly nonlinear effects. This includes kinetic problems that are closely related to higher order moments, especially typical problems with multiple extreme values mathematically:
  (i) Turbulent transport; 
  (ii) Multiscale evolution in burning plasmas ($\alpha$-$i$-$e$ multi-component interaction); 
  (iii) Problems with large gradients (results of stronger magnetic fields and smaller volumes), such as Tokamak pedestal and field reversed configuration (FRC) plasmas; 
  (iv) Magnetic trap configuration with zero magnetic field structure (fusion process of magnetic reconnection in FRC and spherical Tokamak); 
  (v) High frequency injection of fuels into the plasmas; 
  (vi) The cumulative distribution of $\alpha$ particles in fusion plasmas stemming from a pulsating combustion mode. 
These constitute important research frontiers in basic plasma physics theory and non-equilibrium statistical physics, as well as potential key physics issues in the future fusion plasmas and new combustion modes. KMCM is proposed to handle these scenarios which might be characterized by high nonlinearity. 
}

\subsubsection{Some advantages and challenges of KMCM}
\label{Some advantages and challenges of KMCM}

SHE together with KFE framework has various advantages and disadvantages, which are outlined in reference\cite{wang2024Relaxationmodel} and a topic review\cite{wang2024higherorder}. As a moment approach, KMCM possesses its diverse advantages. 

$\blacksquare$ Some advantages of KMCM
\label{Some advantages of KMCM}

\begin{spacing}{0.2}  
\begin{itemize}
    
    \item[(i)] {\textbf{Convergence}: KMCM is derived from the SHE together with KFE framework based on the FDIF hypothesis, accompanied by rigorous mathematical convergence analysis. This is a theoretical physical work, with Wiener's Tauberian theorem \cite{wang2024Relaxationmodel} serving as the foundation to support the fundamental hypothesis.}
    
    \item[(ii)] {\textbf{Theoretical form}: Compared with the traditional moment theory at the same level\cite{Hunana2022}, the arbitrary-order evolution equation (KMEE) within KMCM possesses a unified form, presenting a more standardized and programming-friendly expression. 
    Also because of this,} KMCM is a self-adaptive system by automatically determines the parameters $l_M$, $\bsj_l$ and $\NKa$ (the complexity is determined by the kinetic effects of plasmas).

    {
    \item[(iii)] \textbf{Computability}: For more complex nonlinear Fokker-Planck collision term equations (the collision term part of the 1D-2V VFP equation, while this paper focuses on the non-collision term part), based on the SHE together with KFE frameworks, a meshfree algorithm for the axisymmetric velocity space has been provided in reference \cite{wang2024Relaxationmodel}. The moment approach can adopt the algorithm in the meshfree approach. Moreover, due to the absence of velocity space-related computations, it will enjoy higher computational and storage efficiency theoretically.
    }
    
    \item[(iv)] {\textbf{Efficiency}:} Moreover, the kinetic effects in KMCM are depicted by the higher-order KMEE in the $(j,l)$ space and CPEs in the $(s)$ space. As avoiding the need for detailed information and discrimination of the velocity space, KMCM is proven to be more effective than directly solving the VFP equation for moderately anisotropic plasmas. 
    {For specific problems at specific space-time scales, such as near-equilibrium problem, KMCM can be improved by post-linearization or by specifying larger tolerances.}
    
    \item[(v)] {\textbf{Range of application as a moment model}:} Compared to the traditional moment method, KMCM possesses greater adaptability and a broader application scope. For instance, in strongly non-equilibrium plasmas, the DF might present multi-peak structures or a long tail, which is often challenging to be accurately described by traditional moment methods {based on the perturbation theory}. Conversely, KMCM is able to capture these complex {nonlinear} structures by selecting the appropriate features to maintain high accuracy over a wider range of physical conditions. 
    
    \item[(vi)] {\textbf{Range of application as a kinetic theory}: Just as the conservation laws of mass, momentum and energy are crucial for accurately solving the Vlasov equation, the uniform convergence of higher-order moments is also a property that direct solution methods must fulfill, especially when deal with kinetic problems. As a kinetic moment method, KMCM can naturally achieve the uniform convergence of higher-order moments, similar to the conservation laws.} 
\end{itemize}
\end{spacing}

$\blacksquare$ Some challenges of KMCM
\label{Some challenges of KMCM}

\begin{itemize}
    \item[(i)] {Theoretically,} KMCM implies the utilization of the continuity assumption and is not suitable for situations featuring discontinuities in the phase space. 
    
    \item[(ii)] {Numerically, due to the efficiency of SHE, the current KMCM only has a significant advantage for the moderately anisotropic plasmas.} That is saying that if there are significant asymmetric characteristics in the plasmas system, such as a beam with $\uza/ \vath > 3$ or some group velocity satisfies $\uzhasl/ \vhathsl > 3$, the KMCM given in this study might become inefficient or even fail due to the large value of $l_M$.

    \item[(iii)] {Mathematically,} KMCM is typically a nonlinear system, which requires a sophisticated nonlinear solver, such as optimization algorithm, to solve it with high precision.
    
    \item[(iv)] {Moreover,} how to select the appropriate independent features is another significant issue. This demands not only profound physical perception but also optimization for specific application scenarios.

    \item[(v)] {In efficiency, KMCM is not as efficient as the methods based on perturbation theory for near-equilibrium problem.}

\end{itemize}

Although the outstanding performance of SHE together with KFE framework has been demonstrated in the meshfree approach\cite{wang2024Aconservative} and KMCM has various advantages, solving it still presents a challenge due to its considerable nonlinearity. The key lies in constructing a more ingenious and robust optimization algorithm for solving the CPEs\EQ{CPEs}. This is on our schedule and will be published in the future.

Even so, this new moment theoretical framework based on FDIF hypothesis undoubtedly paves a new path for the study of the Vlasov equation. By integrating physical intuition with mathematical rigor, this framework not only overcomes the drawbacks of traditional moment methods to describe the kinetic behavior of plasmas more accurately, but also lays a solid foundation for future theoretical and applied research. With the continuous enhancement of this framework {in both theory and numerics}, it will play an increasingly significant role in plasma physics, space science, fusion energy, and other related fields.

\section{Conclusion}
\label{Conclusion}

In this paper, we propose a novel transport theory for collisionless fusion plasmas when velocity space exhibits axisymmetry, derived from the 1D-2V Vlasov equation. In this theory, the transport equations, described by kinetic moment evolution equation (KMEE), are obtained from the Vlasov spectrum equation which is given by utilizing the spherical harmonics expansion (SHE) method in the velocity space. Furthermore, the King function expansion (KFE) method is utilized to further reduce the velocity space, which is based on the finitely distinguishable independent features (FDIF) hypothesis. Based on the KFE technology, the characteristic parameter equations (CPEs) are provided, acting as constraint equations of KMEE. Moreover, CPEs also serve as the characteristic closure relation (CCR) of the transport equations in the $(j)$ space. In the $(l)$ space, the natural closure relation (NCR) is employed to enclose the system.

The presented theory, named as kinetic moment-closed model (KMCM), consisting of the transport equations and the closure relations, is typically a nonlinear system. We also precisely prove that the macro state of the collisionless plasmas remains unchanged over time when velocity space exhibits spherical symmetry. Furthermore, the traditional two-fluid equations, as well as conservation laws of mass, momentum and energy, can be derived from the KMCM.
This theory is suitable for moderately anisotropic non-equilibrium plasmas, for which an optimization algorithm is needed for numerical solution. 
In the field of plasmas physics, this nonlinear model might change the solving paradigm of Vlasov equation, and has significant application potential for multi-scale transport problems. At the same time, KMCM can also promote the theoretical development of non-equilibrium statistical physics.

In subsequent studies, we aim to further advance the development of this framework in the following respects. Firstly, the closed theory of the collision aspects also can be directly derived when velocity space exhibits axisymmetry. Secondly, we investigate the scalability of this novel framework for the general velocity space (particularly 3D-3V plasma simulation). Finally, the performance of this framework under various physical conditions needs to be verified through numerical experiments and compared with the traditional approaches.

\section{Acknowledgments}
\label{Acknowledgments}

% I would like to extend my special thanks to Prof. Ge Zhuang for his support. I also  express my gratitude to Hong Qin, Jianyuan Xiao, Xianhao Rao, Zhihui Zou, Pengfei Zhang, Meizhong Ma, Yifeng Zheng, Jian Zheng, Bojing Zhu and Zhe Gao for their valuable contributions to the discussions. I also extend my special thanks to Ye Zhu, Yanxiang Wang, Heyi Li, Shichao Wu, Qiuyue Yuan, Guobing Dong, Guanghui Zhu and Yongli Sun for their supports. This work is also supported by the Strategic Priority Research Program of the Chinese Academy of Sciences (Grant No. XDB0500302).

\begin{CJK*}{UTF8}{gkai} % gbsn 对应宋体，gkai 为楷体，bsmi 为新宋体（繁体）
I would like to extend my special thanks to Prof. Ge Zhuang (庄革) for his support. I also express my gratitude to Hong Qin (秦宏), Jianyuan Xiao (肖建元), Xianhao Rao (饶贤昊), Zhihui Zou (邹志慧), Pengfei Zhang (张鹏飞), Meizhong Ma (马美仲), Yifeng Zheng (郑艺峰), Jian Zheng (郑坚), Bojing Zhu (朱伯靖) and 
% Zhe Gao ({\CJKfontspec{黑体} \selectfont{高喆}})
% Zhe Gao (高 \begin{CJK*}{UTF8}{kai} 喆 \end{CJK*}) 
Zhe Gao (高喆) 
for their valuable contributions to the discussions. Additionally, I will extend my thanks to Ye Zhu (朱叶), Yanxiang Wang (王艳想), Heyi Li (李和意), Shichao Wu (吴士超), Qiuyue Yuan (袁秋月), Guobing Dong (董国兵), Guanghui Zhu (朱光辉) and Yongli Sun (孙永利) for their supports. This work is also supported by the Strategic Priority Research Program of the Chinese Academy of Sciences (Grant No. XDB0500302).
\end{CJK*}

\appendix    % this command starts appendixes

\begin{appendices}

\end{appendices}
 
\end{spacing}
 
%%%%% References %%%%%
\begin{spacing}{0.5}  
 
\
\

% \listoffigures
% \listoftables

\end{spacing}

\bibliographystyle{iopart-num.bst}
\bibliography{Plasma}

\end{document}

%% file: macros.tex
 \global\long\def\calA{\mathcal{A}}
 \global\long\def\calB{\mathcal{B}}
 
 \global\long\def\calD{\mathcal{D}}
 \global\long\def\calE{\mathcal{E}}

 \global\long\def\calK{\mathcal{K}}
 
 \global\long\def\calM{\mathcal{M}}

 \global\long\def\scrT{\mathscr{T}}

 \global\long\def\scrf{\mathscr{f}}

 \global\long\def\scrfh{\hat{\mathscr{f}}}

 \global\long\def\fraka{\mathfrak{a}}

 \global\long\def\rmY{\mathrm{Y}}

 \global\long\def\rmd{\mathrm{d}}
 \global\long\def\rme{\mathrm{e}}

 \global\long\def\rmv{v}

 \global\long\def\rmvh{\hat{v}}

 \global\long\def\bsB{\boldsymbol{B}}
 \global\long\def\bsC{\boldsymbol{C}}
 
 \global\long\def\bsE{\boldsymbol{E}}

 \global\long\def\bsI{\boldsymbol{I}}
 \global\long\def\bsJ{\boldsymbol{J}}

 \global\long\def\bsQ{\boldsymbol{Q}}

 \global\long\def\bse{\boldsymbol{e}}
 \global\long\def\bsf{\boldsymbol{f}}

 \global\long\def\bsj{\boldsymbol{j}}

 \global\long\def\bsq{\boldsymbol{q}}
 \global\long\def\bsr{\boldsymbol{r}}

 \global\long\def\bsu{\boldsymbol{u}}
 \global\long\def\bsv{\boldsymbol{v}}
 \global\long\def\bsw{\boldsymbol{w}}

 \global\long\def\bbN{\mathbb{N}}

 \global\long\def\bbZ{\mathbb{Z}}

 \global\long\def\olra{\overleftrightarrow}

 \global\long\def\olraI{\olra{I}}

 \global\long\def\nh{\hat{n}}
 \global\long\def\uh{\hat{u}}

 \global\long\def\v{\bsv}

 \global\long\def\r{\bsr}

 \global\long\def\olra{\overleftrightarrow}
\global\long\def\olraI{\overleftrightarrow{I}}

 \global\long\def\Ylm{\rmY_l^m}
 \global\long\def\Nlm{N_l^m}
 \global\long\def\Ylmabs{\rmY_l^{\left|m \right|}}

 \global\long\def\max{\mathrm{max}}

 \global\long\def\DDt{\frac{\rmd}{\rmd t}}
 \global\long\def\ddt{\frac{\partial}{\partial t}}

 \global\long\def\ddbfv{\nabla_{\v}}

 \global\long\def\ddrmv{\frac{\partial}{\partial \rmv}}

 \global\long\def\sumloq{\sum_{l=0}^{\infty} }

 \global\long\def\sumlnl{\sum_{m=-l}^l }

  \global\long\def\ddx{\frac{\partial}{\partial x}} 
  \global\long\def\ddz{\frac{\partial}{\partial z}}
  \global\long\def\ddy{\frac{\partial}{\partial y}}

 \global\long\def\e{\rme}
 
 \global\long\def\IFI{\mathrm{_1 F_1}}
 
 \global\long\def\navth{\frac{n_a}{\vath^3}}

 \global\long\def\fl{f_l}

 \global\long\def\flo{f_l^0}

 \global\long\def\flm{f{_{l}^m}}

 \global\long\def\fln{f{_{l-1}}}
 \global\long\def\flp{f{_{l+1}}}

 \global\long\def\fln{f{_{l-1}}}
 \global\long\def\flp{f{_{l+1}}}

 \global\long\def\fln{f{_{l-1}}}
 \global\long\def\flp{f{_{l+1}}}

\global\long\def\fh{\hat{f}}

 \global\long\def\fhl{\fh_l}

 \global\long\def\fln{f{_{l-1}}}
 \global\long\def\flp{f{_{l+1}}}

  \global\long\def\calMjpln{{\calM}{_{j+1,l-1}}}
  \global\long\def\calMjplp{{\calM}{_{j+1,l+1}}}

  \global\long\def\calMjnln{{\calM}{_{j-1,l-1}}}
  \global\long\def\calMjnlp{{\calM}{_{j-1,l+1}}}

\global\long\def\calMjl{\calM_{j,l}}

  \global\long\def\calMjplI{{\calM}{_{j+1,l \pm}}}
  \global\long\def\calMjnlI{{\calM}{_{j-1,l \pm}}}

  \global\long\def\calMjo{{\calM}{_{j,0}}}
  \global\long\def\calMoo{{\calM}{_{0,0}}}

 \global\long\def\Kl{\calK_l}

 \global\long\def\scrfI{\scrf{_1}}

 \global\long\def\scrfI0{\scrf{_1^0}}

 \global\long\def\scrfhI{\scrfh{_1}}

 \global\long\def\scrfhI0{\scrfh{_1^0}}

 \global\long\def\Bn{\bsB}
 \global\long\def\En{\bsE}
 \global\long\def\In{\bsI}
 \global\long\def\Jn{\bsJ}
 
 \global\long\def\Qn{\bsQ}

 \global\long\def\qn{\bsq}
 \global\long\def\wn{\bsw}

 \global\long\def\Al{\calA_l}
 \global\long\def\Bl{\calB_l}
 \global\long\def\El{\calE_l}

  \global\long\def\calMzjplI{{\calM{_z}}{_{j+1,l \pm 1}}}

  \global\long\def\calMzjnlI{{\calM{_z}}{_{j-1,l \pm 1}}}

 \global\long\def\CA{C_A}
 \global\long\def\CE{C_E}
 \global\long\def\CEI{C_E^1}
 \global\long\def\CMA{C{_A}}
 
 \global\long\def\CME{C{_E}}
 \global\long\def\CB{C_B}

 \global\long\def\CAln{\CA{_{l-1}}}
 \global\long\def\CAlp{\CA{_{l+1}}}

 \global\long\def\CAln{\CA{_{l-1}}}
 \global\long\def\CAlp{\CA{_{l+1}}}

 \global\long\def\CMAln{\CMA{_{l-1}}}
 \global\long\def\CMAlp{\CMA{_{l+1}}}

 \global\long\def\CMAln{\CMA{_{l-1}}}
 \global\long\def\CMAlp{\CMA{_{l+1}}}
 
 \global\long\def\CMEjln{\CME{_{j,l-1}}}
 \global\long\def\CMEjlp{\CME{_{j,l+1}}}

 \global\long\def\CEln{\CE{_{l-1}}}
 \global\long\def\CElp{\CE{_{l+1}}}

 \global\long\def\CEln{\CE{_{l-1}}}
 \global\long\def\CElp{\CE{_{l+1}}}
 
 \global\long\def\CEIln{\CEI{_{l-1}}}
 \global\long\def\CEIlp{\CEI{_{l+1}}}

 \global\long\def\CEIln{\CEI{_{l-1}}}
 \global\long\def\CEIlp{\CEI{_{l+1}}}
  
 \global\long\def\CMEln{\CME{_{j,l-1}}}
 \global\long\def\CMElp{\CME{_{j,l+1}}}

 \global\long\def\CMEln{\CME{_{l-1}}}
 \global\long\def\CMElp{\CME{_{l+1}}}

 \global\long\def\CAln{\CA{_{l-1}}}
 \global\long\def\CAlp{\CA{_{l+1}}}

 \global\long\def\CEln{\CE{_{l-1}}}
 \global\long\def\CElp{\CE{_{l+1}}}

 \global\long\def\CEln{\CE{_{l-1}}}
 \global\long\def\CElp{\CE{_{l+1}}}

 \global\long\def\CEIln{\CEI{_{l-1}}}
 \global\long\def\CEIlp{\CEI{_{l+1}}}

 \global\long\def\CEIln{\CEI{_{l-1}}}
 \global\long\def\CEIlp{\CEI{_{l+1}}}

 \global\long\def\CMEln{\CME{_{j,l-1}}}
 \global\long\def\CMElp{\CME{_{j,l+1}}}

 \global\long\def\CBl{\CB{_{l}}}

 \global\long\def\un{\bsu}
 \global\long\def\ua{\un_a}

\global\long\def\uza{u_a}

\global\long\def\uzha{\uh_a}

\global\long\def\uzhasl{\uh_{a_{s,l}}}

 \global\long\def\Ia{\In_a}

\global\long\def\Iza{I_a}

 \global\long\def\vath{\rmv_{ath}}

 \global\long\def\vhathsl{\rmvh_{{ath}_{s,l}}}

 \global\long\def\vhathsl{\rmvh_{{ath}_{s,l}}}

 \global\long\def\xiasl{\xi_{{s,l}}}
\global\long\def\nhasl{\nh_{a_{s,l}}}

\global\long\def\NKa{N_{K_a}}

 \global\long\def\FIG#1{~\ref{#1}}
 \global\long\def\EQ#1{~(\ref{#1})}
 \global\long\def\EQo#1{(\ref{#1})}

 \global\long\def\SEC#1{~\ref{#1}}

%% file: macro_VFP.tex
 \global\long\def\Dr3{\nabla_3}
 \global\long\def\E3{\bsE_3}
 \global\long\def\B3{\bsB_3}

%% file: macro_VFP2.tex
 \global\long\def\fAl{f_{A}{_l}}
 \global\long\def\fEl{f_{E}{_l}}
 \global\long\def\fBl{f_{B}{_l}}

 \global\long\def\CvMAzjl{\bsC{_A^z}{_{j,l}}}

 \global\long\def\CvMEzjl{\bsC{_E^z}{_{j,l}}}